\documentclass[reprint,amsmath,amssymb,aps]{revtex4-1}
\usepackage{amsmath}
\usepackage{amssymb}
\usepackage{textcomp}
\usepackage{graphicx}
\usepackage{dcolumn}
\usepackage{bm}
\DeclareMathOperator{\Tr}{Tr}
\DeclareMathOperator{\RE}{Re}
\DeclareMathOperator{\IM}{Im}
\DeclareMathOperator{\DIV}{div}
\DeclareMathOperator{\ROT}{rot}
\DeclareMathOperator{\sgn}{sgn}

\begin{document}

\preprint{APS/123-QED}

\title{\textbf{Surface plasmon polaritons in planar graphene superlattices}}

\author{Pavel V. Ratnikov}
\email{ratnikov@lpi.ru}
\affiliation{A. M. Prokhorov General Physics Institute, Russian Academy of Sciences, ul. Vavilova 38, 119991 Moscow, Russia}

\date{Received 7 May 2019; revised manuscript received 21 February 2020; accepted 25 February 2020; published 9 March 2020}

\begin{abstract}
Surface plasmon polaritons in planar graphene superlattices with one-dimensional periodic mo-dulation of the band gap were studied. The interminiband contribution to the optical conductivity of this system was found by the equation of motion method for two cases: the Fermi level falls within one of the minigaps and the Fermi
level is located within one of the minibands. It was shown that the optical conductivity of the system varies significantly in these cases. The spectra of surface plasmon polaritons in the system differs for them.

~

\hspace{-0.35cm}\textbf{DOI}: \href{https://doi.org/10.1103/PhysRevB.101.125301}{10.1103/PhysRevB.101.125301}
\end{abstract}



\maketitle

\section{\label{s1}Introduction}

Plasmonics has become a rapidly growing field of solid state physics over the past two decades. In addition to fundamental physics, plasmonics covers a wide range of applications, such as integrated optical circuits \cite{Davis, Heck}, transformation and Fourier optics \cite{Vakil, Wen}, nanophotonics \cite{Xia, LK}, photovoltaics \cite{Mubeen, ARP}, single-molecule detection \cite{Punj}, radiation guiding \cite{HB, DMW}, etc. Most of these applications rely on surface plasmon polaritons (SPPs). SPPs are evanescent electromagnetic waves coupled to the collective plasma oscillations (plasmons), propagating along the surface of a conductor.

The initial studies concerning electromagnetic properties of metal-dielectric boundaries go back to the works by Mi \cite{Mie}, Fano \cite{Fano}, and Ritchie \cite{Rit} for small spherical metallic particles and flat interfaces, respectively. SPPs at a metallic surface has been intensively investigated both in light of the fundamental physics and applications \cite{Zay}. The optical properties of metal nanoparticles show enormous differences with respect to their bulk or thin-film optical responses. While the film absorbs light in all near-infrared and visible regions due to the free-electron absorption, for nanoparticles this process is strongly limited for energies below a given value \cite{Mul}.

The attractiveness of plasmonics is primarily that it is possible with the help of plasmons to concentrate electromagnetic energy at small scales (in comparison with the wavelength of light). Possessing a giant dipole moment, plasmons on these scales play the role of effective intermediaries in the interaction of materials with light. In addition, the properties of plasmons can be controlled within extremely wide limits \cite{Tikh}.

One of the main ways to control plasmon is the design of polariton crystals. Polariton crystals are artificial periodic media, in which along with photon resonances (arising from periodic modulation of the dielectric constant) there are also optically active electron resonances. The first polariton crystals used the Bragg superlattices (SLs) of semiconductor quantum wells (QWs) \cite{Ivch, Koch}. In this case, the role of electron resonances was played by excitons in QWs. Exciton-polariton crystals were later proposed as the photonic crystal slabs, which are planar waveguide layers modulated by one-dimensional (1D) or two-dimensional (2D) gratings of depressions filled with a layered semiconductor with strong exciton resonances \cite{Fuj, Yab, Shi}.

However, the most interesting were the polariton effects in modulated metal-dielectric structures. The surface plasmons play here the role of electron resonances. In fact, the first samples of such ``polariton crystal slabs'' were diffraction gratings. The Wood resonant anomalies \cite{Wood} in the optical spectra of the gratings on the metal surface were first explained by the excitation of surface plasmons in Fano's work \cite{Fano}.

An interest in such structures was subsequently caused by the detection of the extraordinary optical transmission through sub-wavelength hole arrays in a metal layer \cite{Ebb}. The formation of plasmon-waveguide polaritons in arrays of metallic nanoclusters or nanowires on the surface of a planar dielectric waveguide was also found \cite{Lin, Chri}, as well as plasmon effects in metal layers with pore arrays \cite{Tep1, Tep2}.

With the discovery of 2D carbon material graphene \cite{Nov}, new fundamental approaches and technological opportunities have become available in recent years. Graphene is considered to be a promising material for 2D nanoelectronics \cite{RS}. In plasmonics, it can be operating in the mid-infrared and terahertz frequency ranges \cite{Ooi, Guo}. Compared to SPPs in noble metals, SPPs in graphene show stronger mode confinement and relatively greater distance of propagation \cite{Ding, PLCL, Goy}. Graphene has also an attractive property of electrical or chemical tuning \cite{Nov, Pan}.

A frequency of the surface plasmons in doped graphene is proportional to the $\hspace{-0.05cm}^1\hspace{-0.08cm}/\hspace{-0.065cm}_4$power of the charge carriers density, a feature of single-layer graphene, and the $\hspace{-0.05cm}^1\hspace{-0.08cm}/\hspace{-0.065cm}_2$power of the wave number as in 2D electron gas \cite{Jab, Sta}. The latter ceases to be true for plasmons in planar graphene SLs due to the modification of the Coulomb interaction: The plasmon frequency becomes linear in the wave number nearly in the whole plasmon band \cite{Rat1}.

The planar graphene SLs can be formed by alternating strips of gapless graphene and of its gapped modifications \cite{Rat2}. These modifications explore the main property of graphene, namely, its 2D nature. For this, there exist two possible ways: (i) choosing the material of the substrate, e.g., hexagonal boron nitride (hBN) \cite{Gio} on which graphene is deposited, and (ii) depositing atoms or molecules, e.g., hydrogen atoms \cite{Eli} or CrO$_3$ molecules~\cite{Zan} on the surface of a graphene sheet, although the former way manifests dependence on the method of applying a graphene sheet to the substrate and gives small resulting band gap ($\lesssim100$ meV). The Moir\'{e} structure, arising from the lattice mismatching between graphene and substrate, leads to the formation of the secondary Dirac points in the energy spectrum of graphene \cite{Wal, Woo}. In addition, in graphene/hBN heterostructures, the existence of specific collective excitations such as surface plasmon-phonon polaritons due to the strong coupling between SPPs and surface phonon polaritons appears possible \cite{Bra}. Nevertheless, we consider the latter way to be technologically more attractive to obtain gapped graphene (with using, for example, the masking techniques).

Several gapped modifications of graphene with the band gap ranging from about 53 meV to 5.4 eV have been already demonstrated \cite{Gio, Eli, Zan}. In principle, it is possible to form regions of them with semiconductor or dielectric properties on a single sheet of graphene, creating planar heterostructures. The use of gapped graphene to create potential barriers opens up additional possibilities for \emph{band gap engineering} in carbon-based materials~\cite{Han}.

An important step in theoretical research of electron properties of planar graphene SLs was Ref. \cite{Maks}, where the conditions for arising the secondary Dirac points in the energy spectrum of such heterostructures were found. The dispersion law and renormalized group velocities around these points were calculated. At some parameters of the system, interface states can exist near the top of the valence miniband.

\begin{figure}[t!]
\includegraphics[width=0.48\textwidth]{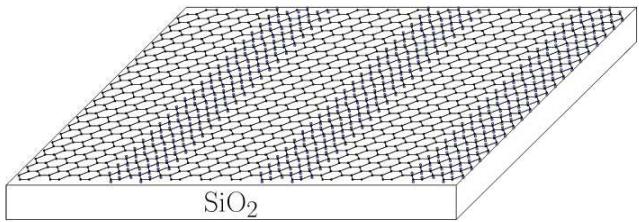}
\caption{\label{f1} An example of a system under consideration: a graphene–graphane SL on a SiO$_2$ substrate (the positions of hydrogen atoms are shown by blue circles).}
\end{figure}

In this paper, we consider a problem of the dispersion relation for SPPs in the planar graphene SLs with 1D periodic modulation of the band gap (one version of such a SL is shown in Fig. \ref{f1}). A few years earlier, SPPs in graphene were discussed in some detail in Ref. \cite{BFPV}. Among other things, the electromagnetic radiation coupling to graphene with 1D periodic modulation of conductivity was considered. The standard approach was used when electric and magnetic fields satisfy the Bloch theorem and they can be written in the form of Fourier-Floquet series. In our case, we proceed from the fact that there are minibands in the energy spectrum of the planar graphene SL (the optical conductivity is calculated as for 2D semiconductors with such energy spectrum), and the fields are also represented in the form of Fourier-Floquet series.

The paper is organized as follows. A model for the planar graphene SLs is presented in Sec.~\ref{s2}. An effective description of charge carriers in these SLs is introduced in Sec.~\ref{s3}. The optical conductivity of the system is analyzed in Sec.~\ref{s4}. The dispersion relation for SPPs is obtained in Sec.~\ref{s5}. The estimation of losses at excitation of these plasmons is given in Sec.~\ref{s6}. Finally, the results of the work are summarised and briefly discussed in Sec.~\ref{s7}.

\section{\label{s2}Model description of the planar graphene SL}

The main concepts concerning the planar SLs based on gapless graphene and on its gapped modifications were reported in Ref. \cite{Rat2}. In this section, we revisit some fundamentals of the model description of charge carriers in these heterostructures.

\begin{figure}[t!]
\includegraphics[width=0.48\textwidth]{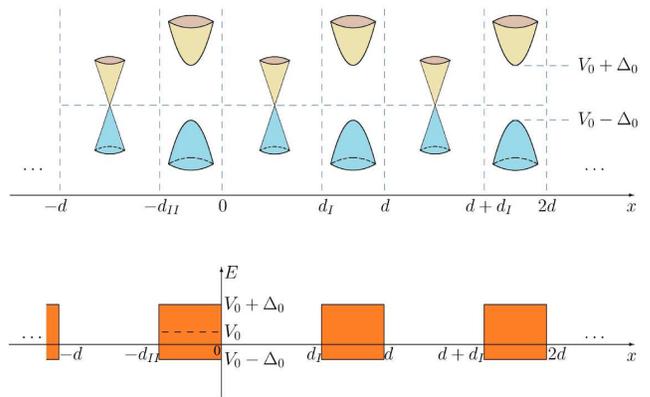}
\caption{\label{f2} Periodic alternating stripes of gapless graphene (regions I) and its gapped modifications (regions II) leads to periodic alternating of the band gap in the space along the $x$ axis. The band gaps in areas II are potential barriers (they are highlighted in orange on the bottom panel), which form 1D periodic Kronig-Penney potential of SLs.}
\end{figure}

Let $x$ and $y$ axes be, respectively, normal and parallel to the interfaces between gapless and gapped graphenes. As in a single graphene sheet, the SL electronic structure is determined by a low-energy dynamics of charge carriers in the vicinity of the Dirac points of the Brillouin zone (BZ). Mathematically, the carriers are described by the envelope wave function $\Psi(x,\,y)$ obeying the Dirac equation in 2D space,
\begin{equation}\label{1}
\left[v_F{\boldsymbol\sigma}\widehat{\bf p}+\sigma_z\Delta(x)+V(x)\right]\Psi(x,\,y)=E\Psi(x,\,y),
\end{equation}
where $v_F\approx10^8$ cm/s is the Fermi velocity, ${\boldsymbol\sigma}=(\sigma_x,\,\sigma_y)$ and $\sigma_z$ are the Pauli matrices, and $\widehat{\bf p}=–i{\boldsymbol\nabla}$ is the momentum operator (here and below $\hbar=1$). The half width of the band gap is a periodic piecewise constant function
\begin{equation*}
\Delta(x)=\begin{cases}0,& d(n-1)<x<-d_\text{II}+dn,\\ \Delta_0,& -d_\text{II}+dn<x<dn,\end{cases}
\end{equation*}
where $n$ is an integer enumerating the supercells, $d_\text{I}$ and $d_\text{II}$ are the widths of strips of the gapless and gapped graphenes, respectively, and $d=d_\text{I}+d_\text{II}$ is the SL period, i.e., the size of the supercell along the $x$ axis (see Fig. \ref{f2}).

The periodic scalar potential $V=V(x)$ can appear due to the difference between the energy positions of the middle of the band gap of the gapped graphene and the Dirac points of
BZ for gapless graphene
\begin{equation*}
V(x)=\begin{cases}0,& d(n-1)<x<-d_\text{II}+dn,\\ V_0,& -d_\text{II}+dn<x<dn.\end{cases}
\end{equation*}
To avoid the production of electron-hole pairs, SL is the first type and the inequality  $|V_0|\leqslant\Delta_0$ must be satisfied.

In general case, the Fermi velocity can differ in graphene modifications. We neglect here the dependence $v_F$ on $x$. We have previously considered SL with alternating Fermi velocity in Ref. \cite{Rat3}.

Since a free motion of charge carriers is realized along the $y$ axis, the solution of Eq. (\ref{1}) for the first supercell has the form
\begin{equation*}
\Psi(x,\,y)=\psi(x)e^{ik_yy},\hspace{0.25cm}0<x<d,
\end{equation*}
where the wave function $\psi(x)$ is a two-component spinor:
\begin{equation*}
\psi(x)=\begin{pmatrix}\psi_u(x)\\ \psi_l(x)\end{pmatrix}.
\end{equation*}

For the $n$th supercell, in view of the periodicity of the system,
\begin{equation*}
\psi_n(x)=\psi(x+(n-1)d).
\end{equation*}
In the QW region $0<x<d_\text{I}$ (region I), the solution of Eq. (\ref{1}) is a linear combination of two spinors with plane waves
\begin{equation}\label{2}
\psi^{(\text{I})}_n(x)=N\begin{pmatrix}a^{(\text{I})}_n\\ b^{(\text{I})}_n\end{pmatrix}e^{ik_\text{I}x}+N\begin{pmatrix}c^{(\text{I})}_n\\ d^{(\text{I})}_n\end{pmatrix}e^{-ik_\text{I}x},
\end{equation}
where $N$ is a normalization factor.

The substitution of the expression Eq. (\ref{2}) into Eq. (\ref{1}) provides the relation between the lower and upper spinor components
\begin{equation*}
b^{(\text{I})}_n=\lambda_+a^{(\text{I})}_n \hspace{0.15cm}\text{and}\hspace{0.15cm} d^{(\text{I})}_n=-\lambda_-c^{(\text{I})}_n,
\end{equation*}
where
\begin{equation*}
\lambda_\pm=\frac{v_F(k_\text{I}\pm ik_y)}{E}.
\end{equation*}
The relation of the charge carrier energy $E$ with $k_\text{I}$ and $k_y$ has the form
\begin{equation*}
E=\pm v_F\sqrt{k^2_\text{I}+k^2_y}
\end{equation*}
(plus for electrons and minus for holes).

It is convenient to represent Eq. (\ref{2}) in a more compact form \cite{Bar}:
\begin{equation}\label{3}
\begin{split}
\psi^{(\text{I})}_n(x)&={\boldsymbol\Omega}_\text{I}(x)\begin{pmatrix}a^{(\text{I})}_n\\ c^{(\text{I})}_n\end{pmatrix},\\
\Omega_\text{I}(x)&=N\begin{pmatrix}1&1\\
\lambda_+&-\lambda_-\end{pmatrix}e^{ik_\text{I}x\sigma_z}.
\end{split}
\end{equation}

When the inequality
\begin{equation}\label{4}
\Delta^2_0+v^2_Fk^2_y-(E-V_0)^2\geqslant0
\end{equation}
is satisfied, the solution of Eq. (\ref{1}) in the barrier region $d_\text{I}<x<d$ (region II) is a linear combination of two spinors with increasing and damped exponents and it can be rewritten in the form analogous to the expression Eqs.~(\ref{3}) (with an accuracy to the substitution $k_\text{I}\rightarrow ik_\text{II}$),
\begin{equation}\label{5}
\begin{split}
\psi^{(\text{II})}_n(x)&={\boldsymbol\Omega}_\text{II}(x)\begin{pmatrix}a^{(\text{II})}_n\\ c^{(\text{II})}_n\end{pmatrix},\\
{\boldsymbol\Omega}_\text{II}(x)&=N\begin{pmatrix}1&1\\ \widetilde{\lambda}_+&-\widetilde{\lambda}_-\end{pmatrix}e^{-k_\text{II}x\sigma_z},
\end{split}
\end{equation}
where
\begin{equation*}
\widetilde{\lambda}_\pm=\frac{iv_F(k_\text{II}\pm k_y)}{E+\Delta_0-V_0},\hspace{0.15cm}k_\text{II}=\frac{1}{v_F}\sqrt{\Delta^2_0+v^2_Fk^2_y-(E-V_0)^2}.
\end{equation*}

When the condition Eq. (\ref{4}) is not satisfied, the solution of Eq. (\ref{1}) in the barrier region becomes oscillating.

The dispersion relation is derived using the transfer matrix method. The transfer matrix \textbf{T} relates the
spinor components for the $n$th supercell to the spinor components of the solution of the same type for the
$(n+1)$th supercell. For example, for the solution in the QW region:
\begin{equation}\label{6}
\begin{pmatrix}a^{(\text{I})}_{n+1}\\ c^{(\text{I})}_{n+1}\end{pmatrix}=\textbf{T}\begin{pmatrix}a^{(\text{I})}_n\\ c^{(\text{I})}_n\end{pmatrix}.
\end{equation}

To determine the \textbf{T} matrix, we use the following boundary conditions:
\begin{equation}\label{7}
\begin{split}
\psi^{(\text{I})}_n(d_{\text{I}-})&=\psi^{(\text{II})}_n(d_{\text{I}+}),\\
\psi^{(\text{II})}_n(d_-)&=\psi^{(\text{I})}_{n+1}(0_+),
\end{split}
\end{equation}
which express the continuity of the solution of the Dirac Eq. (\ref{1}).

The boundary conditions Eqs. (\ref{7}) provide the equalities
\begin{equation*}
\begin{split}
\begin{pmatrix}a^{(\text{II})}_n\\ c^{(\text{II})}_n\end{pmatrix}&={\boldsymbol\Omega}^{-1}_\text{II}(d_\text{I}){\boldsymbol\Omega}_\text{I}(d_I)\begin{pmatrix}a^{(\text{I})}_n\\ c^{(\text{I})}_n\end{pmatrix},\\
\begin{pmatrix}a^{(\text{I})}_{n+1}\\ c^{(\text{I})}_{n+1}\end{pmatrix}&={\boldsymbol\Omega}^{-1}_\text{I}(0){\boldsymbol\Omega}_\text{II}(d)\begin{pmatrix}a^{(\text{II})}_n\\ c^{(\text{II})}_n\end{pmatrix}.
\end{split}
\end{equation*}
According to definition Eq. (\ref{6}) and the last two equalities, we determine the transfer matrix as
\begin{equation}\label{8}
\textbf{T}={\boldsymbol\Omega}^{-1}_\text{I}(0){\boldsymbol\Omega}_\text{II}(d){\boldsymbol\Omega}^{-1}_\text{II}(d_\text{I}){\boldsymbol\Omega}_\text{I}(d_\text{I}).
\end{equation}
The substitution of expressions for ${\boldsymbol\Omega}_\text{I}$ from Eqs. (\ref{3}) and ${\boldsymbol\Omega}_\text{II}$ from Eqs. (\ref{5}) with the corresponding arguments into Eq. (\ref{8}) yields the expressions for elements of transfer matrix
\begin{equation}\label{9}
\begin{split}
T_{11}&=\alpha e^{ik_\text{I}d_\text{I}}\left[(\lambda_-+\widetilde{\lambda}_+)(\lambda_++\widetilde{\lambda}_-)e^{-k_\text{II}d_\text{II}}\right.\\
&\left.-(\lambda_--\widetilde{\lambda}_-)(\lambda_+-\widetilde{\lambda}_+)e^{k_\text{II}d_\text{II}}\right],\\
T_{12}&=2\alpha e^{-ik_\text{I}d_\text{I}}(\lambda_-+\widetilde{\lambda}_+)(\lambda_--\widetilde{\lambda}_-)\sinh(k_\text{II}d_\text{II}),\\ T_{21}&=T_{12}^*,\hspace{0.25cm}T_{22}=T_{11}^*,
\end{split}
\end{equation}
where
\begin{equation*}
\alpha=\frac{1}{(\lambda_++\lambda_-)(\widetilde{\lambda}_++\widetilde{\lambda}_-)}.
\end{equation*}
It is easy to see that $\det\textbf{T}=1$ \cite{McK}.

The dispersion relation is obtained in the form \cite{Rat2, Bar}
\begin{equation}\label{10}
\Tr\textbf{T}=2\cos(k_xd),
\end{equation}
where $k_x$ is the $x$-component of the Bloch wave vector, $k_x\in[-\pi/d,\,\pi/d]$.

Dispersion relation Eq. (\ref{10}) under condition Eq. (\ref{4}) gives the equation \cite{Rat2}
\begin{equation}\label{11}
\begin{split}
&\frac{v^2_Fk^2_\text{II}-v^2_Fk^2_\text{I}+V^2_0-\Delta^2_0}{2v^2_Fk_\text{I}k_\text{II}}\sin(k_\text{I}d_\text{I})\sinh(k_\text{II}d_\text{II})\\ &+\cos(k_\text{I}d_\text{I})\cosh(k_\text{II}d_\text{II})=\cos(k_xd),
\end{split}
\end{equation}
where $k_\text{I}$ and $k_\text{II}$ are implicit functions of $k_x$ and $k_y$.

The passage to the single-band limit is performed in two ways: First, $V_0=\Delta_0$ (QWs only for electrons) and, second, $V_0=-\Delta_0$ (QWs only for holes). The result of the passage coincides with the known nonrelativistic dispersion relation (see, e.g., Ref. \cite{Her}), although the expressions for $k_\text{I}$, $k_\text{II}$, and $E$ are different.

For \emph{Tamm} minibands \cite{Tikh1, Tikh2}, the change $k_\text{I}\rightarrow i\kappa_\text{I}$ should be made in Eq. (\ref{11}):
\begin{equation}\label{12}
\begin{split}
&\frac{v^2_Fk^2_\text{II}+v^2_F\kappa^2_\text{I}+V^2_0-\Delta^2_0}{2v^2_F\kappa_\text{I}k_\text{II}}\sinh(\kappa_\text{I}d_\text{I})\sinh(k_\text{II}d_\text{II})\\ &+\cosh(\kappa_\text{I}d_\text{I})\cosh(k_\text{II}d_\text{II})=\cos(k_xd).
\end{split}
\end{equation}
Equation (\ref{12}) has the solution when $v^2_Fk^2_\text{II}+v^2_F\kappa^2_\text{I}+V^2_0-\Delta^2_0<0$. More detail analysis \cite{Maks} showed that \emph{Tamm} minibands can exist under the condition
\begin{equation}\label{13}
v^2_Fk^2_y<\Delta^2_0\left(\frac{\Delta^2_0}{V^2_0}-1\right).
\end{equation}
In case $V_0>0$ \emph{Tamm} minibands can exist only for~holes, and in case $V_0<0$ they can exist only for electrons. Formally, the condition Eq.~(\ref{13}) coincides with the~qualitative criterion for the existence of interface states when intersecting the dispersion curves of adjoining substances~\cite{Kol}.

\section{\label{s3}Effective description of charge carriers}

For the further analytical study, it is difficult to use the exact spectrum of charge carriers determined by finding the numerical solution of Eq. (\ref{11}). We suggest using the effective spectrum as the spectrum of a model 2D narrow-gap semiconductor with boundaries of BZ along $k_x$ axis, $-\pi/d$ and $\pi/d$. Such consideration has been successfully used when we have determined the plasmon dispersion law in the planar graphene SLs \cite{Rat1}.

We should distinguish two cases: (i) the Fermi level falls within one of the minigaps and (ii) the Fermi
level is located within one of the minibands.

In the former case, all minibands lying below the Fermi level are completely occupied and the oscillations
of the electron (hole) density occur only in the direction of the free motion of charge carriers (along
the direction perpendicular to the Kronig-Penney potential of SLs). This is a quasi-1D motion.

In the latter case, the miniband containing the Fermi level is occupied only partially, whereas all
lower bands (if such bands exist) are completely occupied. In the partially occupied miniband, the oscillations
of electron (hole) density can also occur along the Kronig-Penney potential of SLs. This is a quasi-2D motion.

Using the electric field effect in the system under consideration, it is easy to achieve a crossover between quasi-1D and quasi-2D regimes. For simplicity, we consider below the situation with the filling (complete or partial) of only one lowest electron miniband or the highest hole miniband.

At sufficiently large values of $\Delta_0$ and $d_\text{II}$, the minibands are rather narrow (we shall
specify this condition below). For example, the charge carriers energy spectrum in the lowest electron or the highest hole miniband is (plus corresponds to electrons, minus corresponds to holes)
\begin{equation}\label{14}
E\approx V_\text{eff}\pm\sqrt{\Delta^2_\text{eff}+v^2_Fk^2_y}.
\end{equation}
Here, $\Delta_\text{eff}$ and $V_\text{eff}$ play the role of the effective band gap and the effective work function, respectively.

We can write the effective Hamiltonian corresponding to the approximate dispersion law given by Eq. (\ref{14})
as the Dirac Hamiltonian in terms of $2\times2$ matrices:
\begin{equation}\label{15}
\widehat{H}^{(\text{1D})}_\text{eff}=v_F\sigma_y\widehat{p}_y+\sigma_z\Delta_\text{eff}+V_\text{eff}.
\end{equation}

The charge carriers have the effective mass
\begin{equation*}
m^*=\frac{\Delta_\text{eff}}{v^2_F}.
\end{equation*}

Using dispersion relation Eq. (\ref{11}) and assuming that $\left|V_\text{eff}\right|<\Delta_\text{eff}\ll\Delta_0$, we can easily deduce the following estimates for the $m$th miniband ($m=0,\,1,\,2,\,\ldots$) \cite{Rat1}:
\begin{equation}\label{16}
\begin{split}
\Delta_\text{eff}&=\frac{(2m+1)\pi v_F}{2d_\text{I}}\left[1-\frac{v_F}{d_\text{I}\Delta_0}\right],\\
V_\text{eff}&=\frac{v_F}{d_\text{I}\Delta_0}V_0.
\end{split}
\end{equation}

In the case under study, the minibands have an exponentially small width owing to an exponentially
small probability for charge carriers to tunnel through the barriers. In this limit, we obtain the following estimate for the miniband width:
\begin{equation}\label{17}
\delta E=\frac{4v_F}{d_\text{I}}\exp\left(-\frac{d_\text{II}}{v_F}\Delta_0\right).
\end{equation}

The condition defining the narrow minibands is $\delta E\ll\Delta_\text{eff}$. Comparing the expression for $\Delta_\text{eff}$ in Eqs. (\ref{16}) with Eq. (\ref{17}), we find the condition $\Delta_0\gtrsim2v_F/d_\text{II}$.

The Fermi energy $E_F$ is related to the 1D Fermi momentum $p_F$ as follows ($\widetilde{E}_F=E_F-V_\text{eff}$):
\begin{equation*}
|\widetilde{E}_F|=\sqrt{\Delta^2_\text{eff}+v^2_Fp^2_F}.
\end{equation*}
The 1D Fermi momentum is expressed in terms of the charge carrier density $n_\text{2D}$
\begin{equation*}
p_F=\frac{\pi}{g}n_\text{2D}d,
\end{equation*}
where $g=g_sg_v$ is the degeneracy multiplicity: $g_s=2$ and $g_v=2$ are the degeneracy multiplicity by spin and valley, respectively.

In the quasi-2D case, in addition to the free motion along the gapless graphene strips, charge carriers move across the potential barriers. These types of motion occur at different velocities: at $v_\parallel$ for the free motion and at a much lower velocity $v_\perp\ll v_\parallel$ for the motion perpendicular to the strips (since the probability of tunneling through the potential barrier is small). This means the quasi-2D anisotropic motion of charge carriers. The corresponding values of $v_\parallel$ and $v_\perp$ are selected by fitting the approximate dispersion law. For example, the approximate dispersion law in the lowest electron or the highest hole miniband is
\begin{equation}\label{18}
E\approx V_\text{eff}\pm\sqrt{\Delta^2_\text{eff}+v^2_\perp k^2_x+v^2_\parallel k^2_y}.
\end{equation}
Parameters $\Delta_\text{eff}$ and $V_\text{eff}$ play the same role as in the quasi-1D case and the estimates Eqs. (\ref{16}) can be also applied to them under the conditions indicated above. With a good accuracy, we can assume for all minibands $v_\parallel\approx v_F$.

The effective Hamiltonian with eigenvalues Eq. (\ref{18}) has the form
\begin{equation}\label{19}
\widehat{H}^{(\text{2D})}_\text{eff}=v_\perp\sigma_x\widehat{p}_x+v_\parallel\sigma_y\widehat{p}_y+\sigma_z\Delta_\text{eff}+V_\text{eff}.
\end{equation}

The energy spectrum is similar to that of an anisotropic narrow-band semiconductor with the effective
masses
\begin{equation*}
\begin{split}
m^*_\perp&=\Delta_\text{eff}/v^2_\perp,\\
m^*_\parallel&=\Delta_\text{eff}/v^2_\parallel.
\end{split}
\end{equation*}

\section{\label{s4}Optical conductivity of the system}

The optical conductivity of the system is a sum of~two contributions: (i) a Drude contribution describing intraminiband transitions $\sigma^\text{intra}$ and (ii) a term corresponding interminiband processes $\sigma^\text{inter}$.

The value $\sigma^\text{intra}$ is easily found from the kinetic equation in the $\tau$ approximation ($\gamma=\tau^{-1}$ is the inverse relaxation time) \cite{Rat1}:
\begin{itemize}
\item[(i)] in The quasi-1D case:
\begin{equation}\label{20}
\sigma^\text{intra}=\frac{ige^2v^2_Fp_F}{\pi|\widetilde{E}_F|(\omega+i\gamma)},
\end{equation}
\item[(ii)] The quasi-2D case:
\begin{equation}\label{21}
\begin{split}
\sigma^\text{intra}_{xx}=\frac{ige^2}{\pi(\omega+i\gamma)}\frac{\widetilde{E}^2_F-\Delta^2_\text{eff}}{|\widetilde{E}_F|}\frac{v_\perp}{v_\parallel},\\
\sigma^\text{intra}_{yy}=\frac{ige^2}{\pi(\omega+i\gamma)}\frac{\widetilde{E}^2_F-\Delta^2_\text{eff}}{|\widetilde{E}_F|}\frac{v_\parallel}{v_\perp}.
\end{split}
\end{equation}
\end{itemize}

The values of $v_\perp$ and $v_\parallel$ refer to the partially occupied miniband.

The contribution of interminiband processes to the optical conductivity is calculating by the equation of motion method \cite{Per, Fer}. For definiteness, we consider in detail the quasi-2D case (the quasi-1D case is analogously considered). The formula for the optical conductivity can be written as
\begin{equation}\label{22}
\begin{split}
&\sigma_{ij}(\omega)=\frac{gS}{i\omega}\sum_{m,m^\prime}\sum_{\zeta,\zeta^\prime=\pm1}\sum_{{\bf k},{\bf k}^\prime}\left\langle m,\,\zeta,\,{\bf k}\left|\widehat{J}^{(m^\prime)}_i\right|m^\prime,\,\zeta^\prime,\,{\bf k}^\prime\right\rangle\\
&\left\langle m^\prime,\,\zeta^\prime,\,{\bf k}^\prime\left|\widehat{J}^{(m)}_j\right|m,\,\zeta,\,{\bf k}\right\rangle\frac{n_F[E_{m\zeta}({\bf k})]-n_F[E_{m^\prime\zeta^\prime}({\bf k}^\prime)]}{E_{m^\prime\zeta^\prime}({\bf k}^\prime)-E_{m\zeta}({\bf k})-\omega-i\Gamma},
\end{split}
\end{equation}
where $S$ is the area of the system, $m$ and $m^\prime$ number the minibands ($m,\,m^\prime=0,\,1,\,2,\ldots$), $n_F[E]$ is the Fermi-Dirac distribution function and, for simplicity, we assume $n_F[E]=\theta(E_F-E)$ ($E_F$ is the Fermi energy), $\zeta$ and $\zeta^\prime$ are signs of an energy of the charge carriers ($\zeta,\,\zeta^\prime=+1$ for electrons, and $\zeta,\,\zeta^\prime=-1$ for holes), $E_{m\zeta}({\bf k})=V^{(m)}_\text{eff}+\zeta\varepsilon^{(m)}_{\bf k}$ with $\varepsilon^{(m)}_{\bf k}=\sqrt{\Delta^{(m)2}_\text{eff}+(-1)^mv^{{(m)}2}_\perp k^2_x+v^{{(m)}2}_\parallel k^2_y}$, $\widehat{J}^{(m)}_{i,j}$ and $\widehat{J}^{(m^\prime)}_{i,j}$ are the current density operators ($i,\,j=x,\,y$): $\widehat{J}^{(m)}_x=\frac{e}{S}v^{(m)}_\perp\sigma_x\hspace{0.15cm}\text{and}\hspace{0.15cm}\widehat{J}^{(m)}_y=\frac{e}{S}v^{(m)}_\parallel\sigma_y$. The eigen wave function of the Hamiltonian Eq. (\ref{19}) with parameters $v^{(m)}_\perp$, $v^{(m)}_\parallel$, $\Delta^{(m)}_\text{eff}$, and $V^{(m)}_\text{eff}$ for the $m$th miniband is
\begin{equation*}
\left|m,\,\zeta,\,{\bf k}\right\rangle=\frac{a_{m\zeta{\bf k}}}{\sqrt{2S}}\begin{pmatrix}1\\ b_{m\zeta{\bf k}}\end{pmatrix}e^{i{\bf k}\cdot{\bf r}},
\end{equation*}
where
\begin{equation*}
a_{m\zeta{\bf k}}=\sqrt{1+\zeta\frac{\Delta^{(m)}_\text{eff}}{\varepsilon^{(m)}_{\bf k}}}\hspace{0.15cm}\text{and}\hspace{0.15cm}b_{m\zeta{\bf k}}=\frac{v^{(m)}_\perp k_x+iv^{(m)}_\parallel k_y}{\Delta^{(m)}_\text{eff}+\zeta\varepsilon^{(m)}_{\bf k}}
\end{equation*}
for even $m$ and
\begin{equation*}
\begin{split}
&a_{m\zeta{\bf k}}=\\
&\frac{\left|\Delta^{(m)}_\text{eff}+\zeta\varepsilon^{(m)}_{\bf k}\right|}{\sqrt{\zeta\varepsilon^{(m)}_{\bf k}\left(\Delta^{(m)}_\text{eff}+\zeta\varepsilon^{(m)}_{\bf k}\right)+v^{(m)}_\perp k_x\left(v^{(m)}_\perp k_x+v^{(m)}_\parallel k_y\right)}}
\end{split}
\end{equation*}
and
\begin{equation*}
b_{m\zeta{\bf k}}=\frac{i\left(v^{(m)}_\perp k_x+v^{(m)}_\parallel k_y\right)}{\Delta^{(m)}_\text{eff}+\zeta\varepsilon^{(m)}_{\bf k}}
\end{equation*}
for odd $m$.

Here, we also distinguish the inverse relaxation times $\gamma$ and $\Gamma$ for intraminiband and interminiband transitions, respectively, because these processes are essentially different type ones.

\begin{figure}[t!]
\begin{center}
\includegraphics[width=0.48\textwidth]{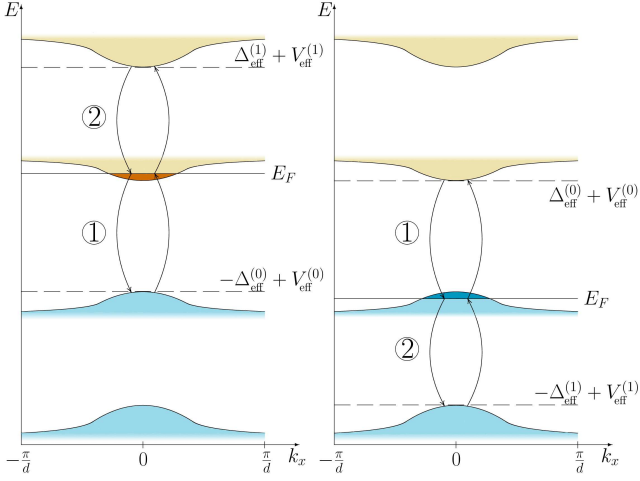}
\caption{\label{f3} Interminiband transitions at low frequencies in the case of partially occupation of the lower electron miniband (the left panel) or upper hole miniband (the right panel). Figures 1 and 2 in the circles correspond to the contributions $\sigma^{(1)}_{ij}$ and $\sigma^{(2)}_{ij}$, respectively. A dependence of the energy $E$ on $k_y$ is qualitatively shown by the color selection.}
\end{center}
\end{figure}

Now, we calculate the interminiband contribution for the case when Fermi level is located within the lower electron miniband or the upper hole miniband. We have two options for interminiband transitions in the formula Eq. (\ref{22}): (1) $m=m^\prime=0$, $\zeta^\prime\neq\zeta$ (transitions between the lower electron miniband and the upper hole miniband, see Fig. \ref{f3}), (2) $m=0$, $m^\prime=1$ or $m=1$, $m^\prime=0$, $\zeta^\prime=\zeta=\sgn(E_F)$ (transitions between the lower electron miniband and the nearest electron miniband or the upper hole miniband and the nearest hole miniband, see Fig. \ref{f3}).

\begin{widetext}
We obtain for the former case
\begin{equation}\label{23}
\begin{split}
\RE\sigma^{(1)}_{xx}(\omega)=&\frac{ge^2}{16}\frac{v^{(0)}_\perp}{v^{(0)}_\parallel}\left(1+\frac{4\Delta^{(0)2}_\text{eff}}{\omega^2+\Gamma^2}\right)\left(1+\frac{1}{\pi}\arctan\frac{\omega-2|\widetilde{E}_F|}{\Gamma}-\frac{1}{\pi}\arctan\frac{\omega+2|\widetilde{E}_F|}{\Gamma}\right),\\
\IM\sigma^{(1)}_{xx}(\omega)=&-\frac{ge^2}{32\pi}\frac{v^{(0)}_\perp}{v^{(0)}_\parallel}\left(1+\frac{4\Delta^{(0)2}_\text{eff}}{\omega^2+\Gamma^2}\right)\ln\frac{(\omega+2|\widetilde{E}_F|)^2+\Gamma^2}{(\omega-2|\widetilde{E}_F|)^2+\Gamma^2}.
\end{split}
\end{equation}

A characteristic logarithm factor appears, as for the imaginary part of the interband contribution to the optical conductivity of graphene (see Ref. \cite{BFPV} and references therein). We also note that the logarithmic divergence at $\omega=2|\widetilde{E}_F|$ in the limit $\Gamma\rightarrow0$ is associated with the Kohn anomaly in graphene.

For simplicity of calculations, we assume also for the latter case $v^{(1)}_\perp\approx v^{(0)}_\perp$ and $v^{(1)}_\parallel\approx v^{(0)}_\parallel$ and the smallness of miniband occupation $\sqrt{\widetilde{E}_F^2-\Delta^{(0)2}_\text{eff}}\ll\Delta^{(0)}_\text{eff}$. Then, we obtain
\begin{equation}\label{24}
\begin{split}
\RE\sigma^{(2)}_{xx}(\omega)\approx&\frac{ge^2\Gamma}{\pi}\frac{v^{(0)}_\perp}{v^{(0)}_\parallel}\frac{\left(\Delta^{(1)}_\text{eff}-\Delta^{(0)}_\text{eff}\right)\left(\widetilde{E}_F^2-\Delta^{(0)2}_\text{eff}\right)}
{\left[(\Delta^{(1)}_\text{eff}-\Delta^{(0)}_\text{eff}-\omega)^2+\Gamma^2\right]\left[(\Delta^{(1)}_\text{eff}-\Delta^{(0)}_\text{eff}+\omega)^2+\Gamma^2\right]},\\
\IM\sigma^{(2)}_{xx}(\omega)\approx&-\frac{ge^2}{2\pi\omega}\frac{v^{(0)}_\perp}{v^{(0)}_\parallel}\left(\Delta^{(1)}_\text{eff}-\Delta^{(0)}_\text{eff}\right)\left(\widetilde{E}_F^2-\Delta^{(0)2}_\text{eff}\right)\\
&\times\left\{\frac{(\Delta^{(1)}_\text{eff}-\Delta^{(0)}_\text{eff})^2-\omega^2+\Gamma^2}{\left[(\Delta^{(1)}_\text{eff}-\Delta^{(0)}_\text{eff}-\omega)^2+\Gamma^2\right]\left[(\Delta^{(1)}_\text{eff}-\Delta^{(0)}_\text{eff}+\omega)^2+\Gamma^2\right]}
-\frac{1}{(\Delta^{(1)}_\text{eff}-\Delta^{(0)}_\text{eff})^2+\Gamma^2}\right\},
\end{split}
\end{equation}
where we excluded from $\IM\sigma^{(2)}_{xx}(\omega)$ the term divergent as $1/\omega$ at small $\omega$.

We see that $\sigma^{(2)}_{xx}(\omega)$ is suppressed in comparison with $\sigma^{(1)}_{xx}(\omega)$ owing to the factor $\left(\widetilde{E}_F^2-\Delta^{(0)2}_\text{eff}\right)/\Delta^{(0)2}_\text{eff}$. For other transitions through one or more of the minibands, the situation is analogous: instead of $\Delta^{(1)}_\text{eff}$, there will be $\Delta^{(m)}_\text{eff}$ with $m=2,\,3,\ldots$ and we will have additional numerical smallness due to $\Delta^{(m)}_\text{eff}>\Delta^{(1)}_\text{eff}$ [according to the evaluation Eqs. (\ref{16}) $\Delta^{(m)}_\text{eff}=(2m+1)\Delta^{(0)}_\text{eff}$]. So, we can neglect contributions to the optical conductivity from transitions that are different from transitions between neighboring minibands, one of which is the Fermi level. We have the result for the $xx$ component of the optical conductivity tensor in the quasi-2D case $\sigma^\text{inter}_{xx}(\omega)=\sigma^{(1)}_{xx}(\omega)+\sigma^{(2)}_{xx}(\omega)$.
The answer for $\sigma^\text{inter}_{yy}(\omega)$ differs from $\sigma^\text{inter}_{xx}(\omega)$ by the replacements $v^{(0)}_\perp\rightleftarrows v^{(0)}_\parallel$.

In the quasi-1D case, when the Fermi level falls into the minigap, we have
\begin{equation}\label{25}
\begin{split}
\RE\sigma^{(1)}(\omega)&=\frac{ge^2v_F\Gamma}{2\pi\Delta^{(0)2}_\text{eff}}I\left(\frac{\omega}{2\Delta^{(0)}_\text{eff}},\,\frac{\Gamma}{2\Delta^{(0)}_\text{eff}}\right),\\
\IM\sigma^{(1)}(\omega)&=-\frac{ge^2v_F}{\pi\omega}J\left(\frac{\omega}{2\Delta^{(0)}_\text{eff}},\,\frac{\Gamma}{2\Delta^{(0)}_\text{eff}}\right),
\end{split}
\end{equation}
where we introduced the functions
\begin{equation*}
\begin{split}
I(a,\,b)&=\int_{x_F}^\infty\frac{dx}{\sqrt{x^2-1}\left[(x-a)^2+b^2\right]\left[(x+a)^2+b^2\right]},\\
J(a,\,b)&=\int_{x_F}^\infty\frac{dx}{\sqrt{x^2-1}}\left\{\frac{x^2-a^2+b^2}{\left[(x-a)^2+b^2\right]\left[(x+a)^2+b^2\right]}-\frac{1}{x^2+b^2}\right\}.
\end{split}
\end{equation*}
The notation $x_F=1$ is introduced for the case when the Fermi level falls within the minigap between the lower electron and the upper hole minibands; $x_F=|\widetilde{E}_F|/\Delta^{(0)}_\text{eff}$ for the case when the Fermi level falls within the minigap between the lower and the next electron minibands or between the upper and the next hole minibands, $x_F>1$. In this case, we have also the second type contribution which is easily obtained for the small miniband occupation:
\begin{equation}\label{26}
\begin{split}
\RE\sigma^{(2)}(\omega)=&\frac{4ge^2v_F\Gamma}{\pi}\frac{\left(\Delta^{(1)}_\text{eff}-\Delta^{(0)}_\text{eff}\right)\sqrt{\widetilde{E}^2_F-\Delta^{(0)2}_\text{eff}}}
{\left[(\Delta^{(1)}_\text{eff}-\Delta^{(0)}_\text{eff}-\omega)^2+\Gamma^2\right]\left[(\Delta^{(1)}_\text{eff}-\Delta^{(0)}_\text{eff}+\omega)^2+\Gamma^2\right]},\\
\IM\sigma^{(2)}(\omega)=&-\frac{2ge^2v_F}{\pi\omega}\left(\Delta^{(1)}_\text{eff}-\Delta^{(0)}_\text{eff}\right)\sqrt{\widetilde{E}^2_F-\Delta^{(0)2}_\text{eff}}\\
&\times\left\{\frac{(\Delta^{(1)}_\text{eff}-\Delta^{(0)}_\text{eff})^2-\omega^2+\Gamma^2}{\left[(\Delta^{(1)}_\text{eff}-\Delta^{(0)}_\text{eff}-\omega)^2+\Gamma^2\right]\left[(\Delta^{(1)}_\text{eff}-\Delta^{(0)}_\text{eff}+\omega)^2+\Gamma^2\right]}
-\frac{1}{(\Delta^{(1)}_\text{eff}-\Delta^{(0)}_\text{eff})^2+\Gamma^2}\right\}.
\end{split}
\end{equation}
\end{widetext}

\begin{figure}[b!]
\begin{center}
\includegraphics[width=0.48\textwidth]{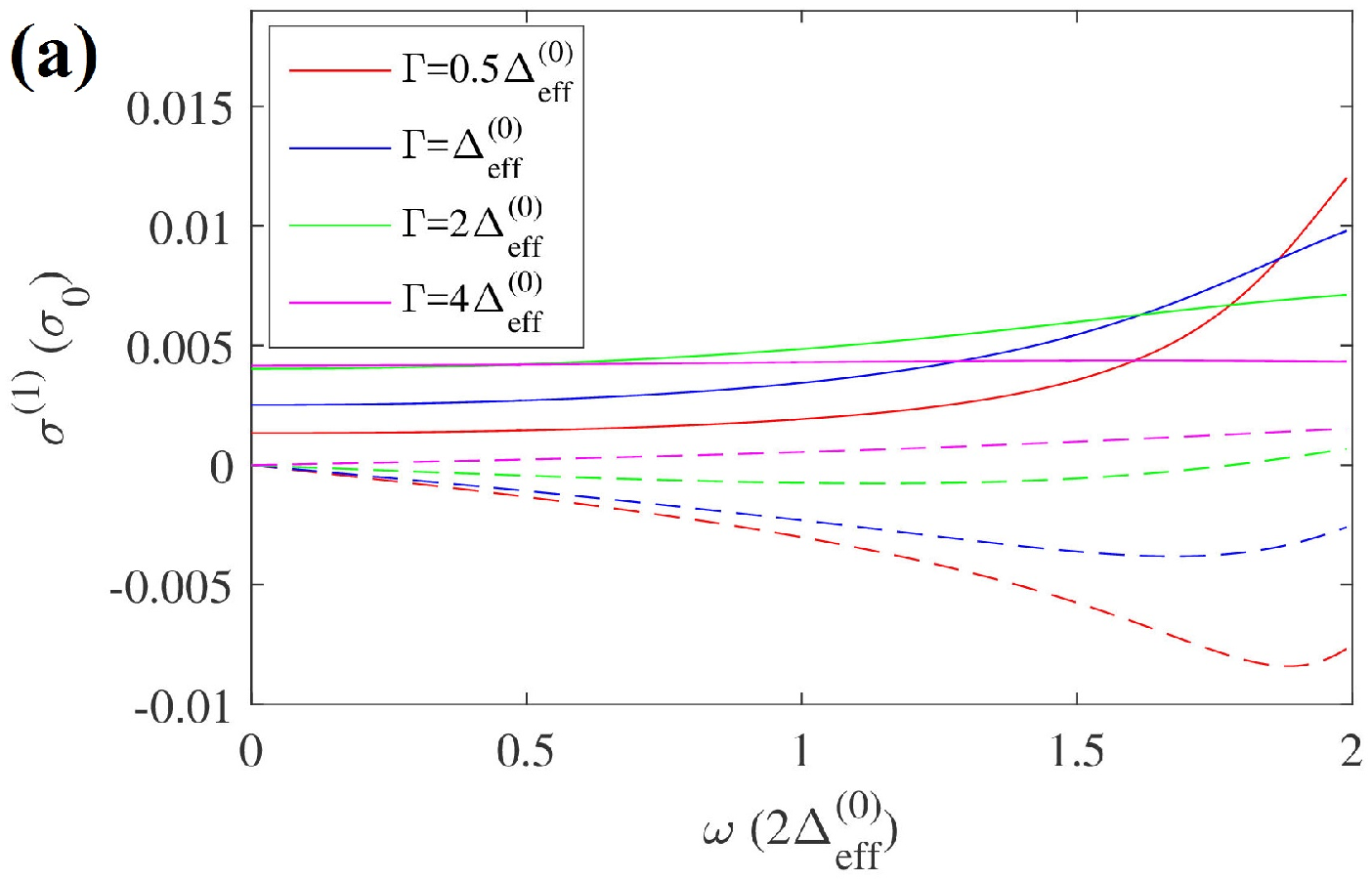}
\includegraphics[width=0.48\textwidth]{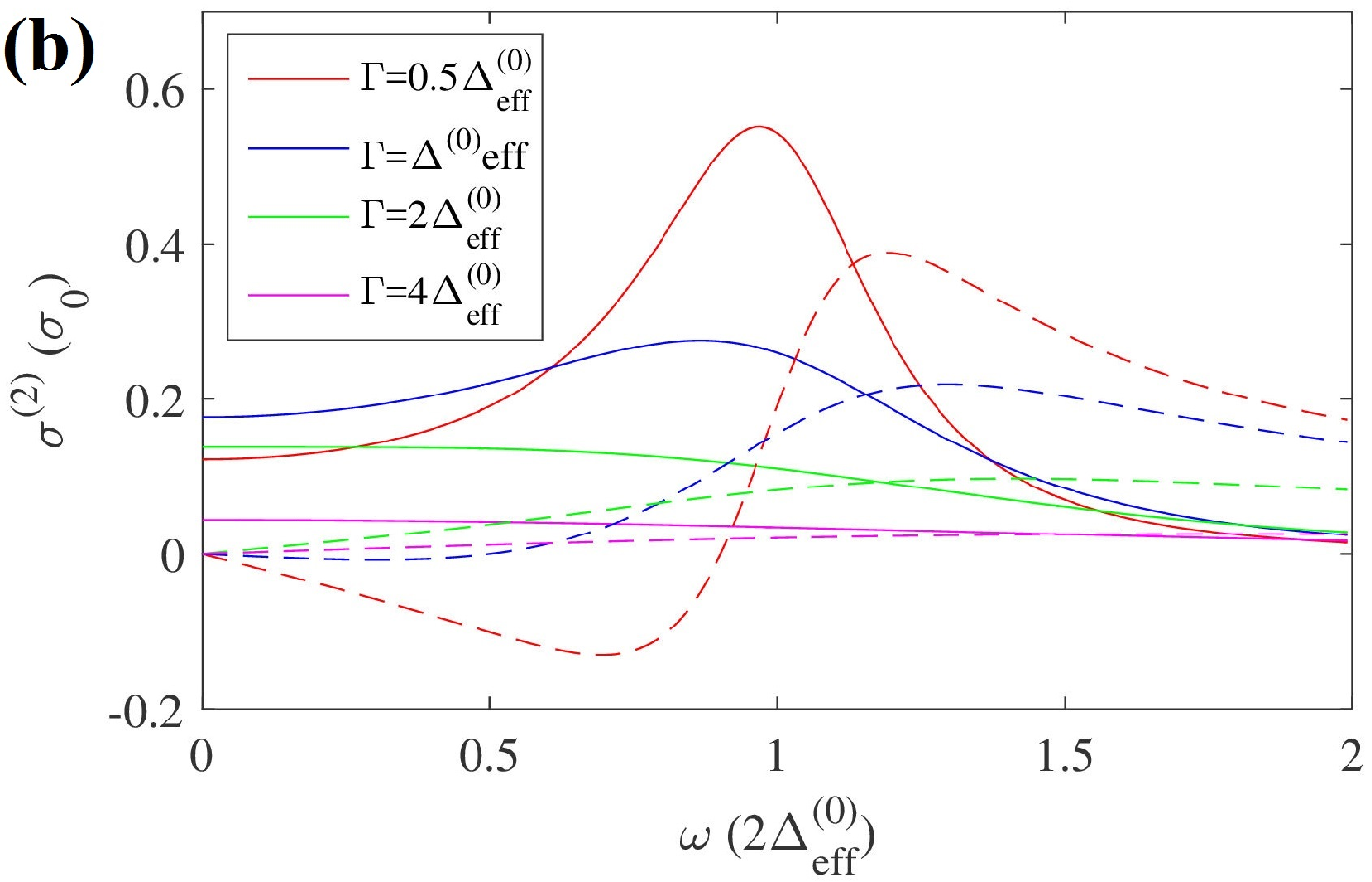}
\includegraphics[width=0.48\textwidth]{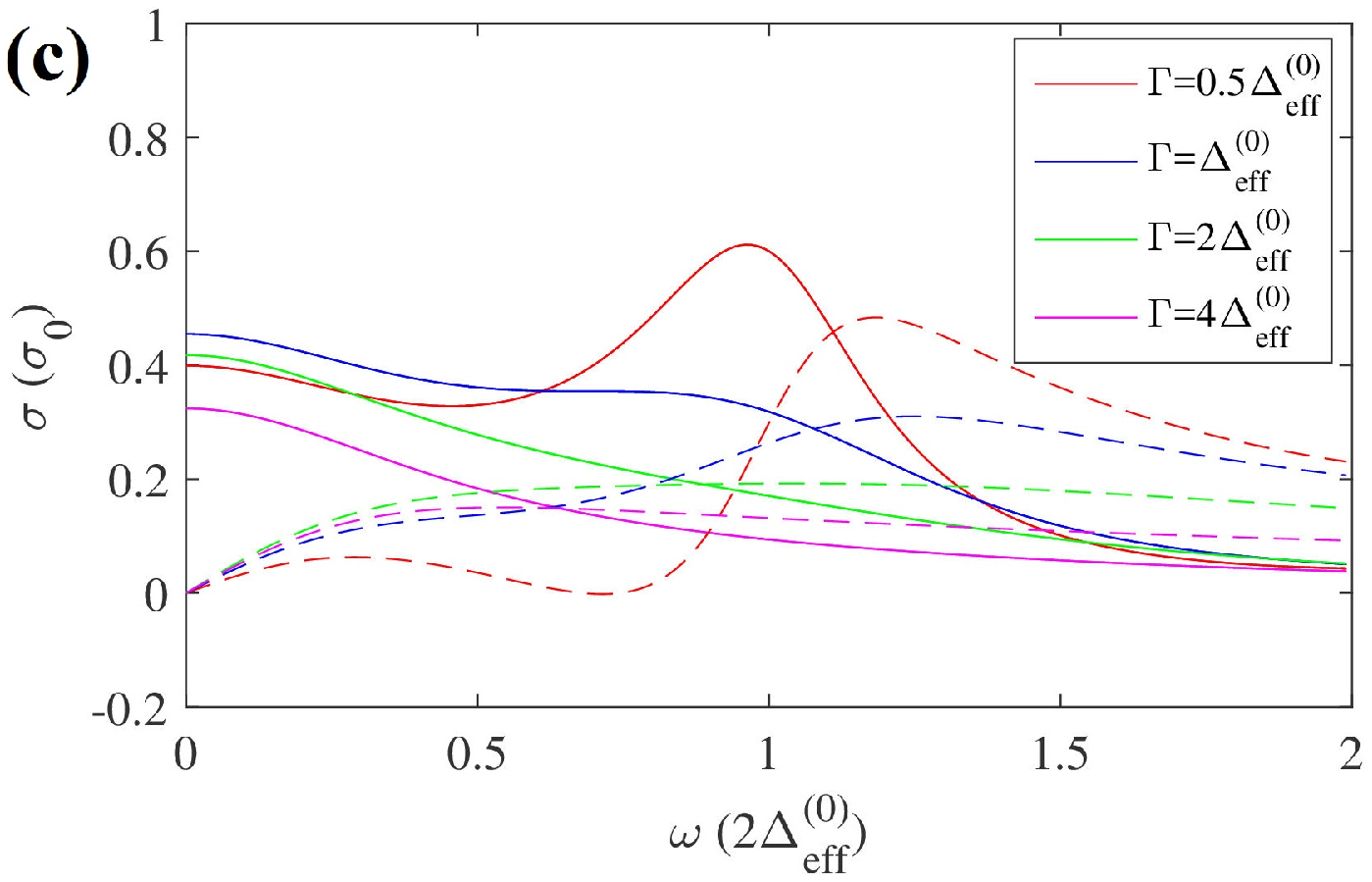}
\caption{\label{f4} Real (solid lines) and imaginary (dashed lines) parts of the optical conductivity. (a) Contribution of the first-type interminiband transitions for four values of $\Gamma$. (b)  Contribution of the second-type interminiband transitions for the same values of $\Gamma$. (c) The sum of these contributions together with the Drude conductivity ($\gamma=\Delta^{(0)}_\text{eff}$).}
\end{center}
\end{figure}

Let us illustrate an interesting feature of the case when the Fermi level lies in the minigap between the lower and the next electron minibands (between the upper and the next hole minibands), when $x_F>1$. We take $x_F=2$. This will ensure the position of the Fermi level approximately in the middle of the minigap [according to the first relation in Eqs. (\ref{16}) $\Delta^{(1)}_\text{eff}=3\Delta^{(0)}_\text{eff}$, and the Fermi level is $\widetilde{E}_F=2\Delta^{(0)}_\text{eff}$]. For generality, we will count the frequency in units of the minigap between the lower electron and upper hole minibands $2\Delta^{(0)}_\text{eff}$, and the contribution to the optical conductivity of interminiband transitions in units of $\sigma_0=ge^2v_F/\Delta^{(0)}_\text{eff}$. Figure \ref{f4} shows numerical calculation with using of the formulas Eqs. (\ref{20}), (\ref{25}), and (\ref{26}). The striking feature in the optical conductivity was that the contribution of the second-type interminiband transitions turned out to be the leading contribution with respect to the first type.

At the end of this section, a few explanations should be made why we neglect nonlocal effects arising from the spatial dispersion of the optical conductivity. We consider graphene from the standpoint of the electrodynamics of continuous media as an infinitely thin conductive film. The inclusion of a spatial dispersion of the optical conductivity has the same role as the allowance for a spatial dispersion of the permittivity. Therefore, the condition when it is possible to neglect the effects associated with the dependence of the optical conductivity on the wave vector $q$ coincides with the condition of neglecting such a dependence for the dielectric function \cite{LL},
\begin{equation}\label{27}
qr_0\ll1,
\end{equation}
where $r_0=\min\left\{v/\omega,\,l\right\}$ is the characteristic distance over which the kernel of the integral expression for the electrical induction is non-zero; $v$ is the mean velocity of charge carriers, and $l$ is their mean free path.

Here, we assume that there is a ballistic regime of passage of charge carriers through many supercells of SLs under consideration, $l\gg d$ and $l>v/\omega$. So, we have $r_0=v/\omega$ with $v=\langle|\mathbf{v}|\rangle$, $\mathbf{v}=\partial E_{m\zeta}({\bf k})/\partial\mathbf{k}$. Averaging is performed within the Fermi sphere:
\begin{equation*}
\langle|\mathbf{v}|\rangle=\frac{\int|\mathbf{v}|n_F[E_{m\zeta}({\bf k})]d^2k}{\int n_F[E_{m\zeta}({\bf k})]d^2k}.
\end{equation*}
It leads to an answer for the quasi-2D case in the form
\begin{equation*}
\begin{split}
v=&\frac{2v_\perp\Delta^2_\text{eff}}{\pi(\widetilde{E}_F^2-\Delta^2_\text{eff})}E\left(1-(v_\perp/v_\parallel)^2\right)\\
&\times\left[\frac{\widetilde{E}_F\sqrt{\widetilde{E}_F^2-\Delta^2_\text{eff}}}{\Delta^2_\text{eff}}
-\ln\frac{\widetilde{E}_F+\sqrt{\widetilde{E}_F^2-\Delta^2_\text{eff}}}{\Delta_\text{eff}}\right],
\end{split}
\end{equation*}
where $E(x)$ is the complete elliptic integral of the second kind \cite{AS}.

To demonstrate the characteristic values, we give a calculation of the mean velocity of charge carriers $v$ for an example of SLs considered below in the end of Sec.~\ref{s5} with $\Delta_0=60$ meV, $d\approx418$ nm, $\Delta^{(0)}_\text{eff}\approx1.58$ meV, $E_F=2$ meV, $v_\perp\approx5.95\times10^7$ cm/s, and $v_\parallel\approx8.5\times10^7$ cm/s. We find $v\approx2.27\times10^7$ cm/s and $r_0\simeq150$ nm at the characteristic value of SPP energy $\omega\simeq1$ meV. The wave vector characteristic value is $q\simeq10^2$ cm$^{-1}$. Thus, we obtain $r_0q\simeq1.5\times10^{-3}$ and the condition Eq. (\ref{27}) is more than fulfilled. Accordingly, there is no need to consider effects due to the possible dependence of conductivity on the wave vector. However, we explain what these effects in general lines are.

We note immediately that for the physically meaningful discussion of these effects, it is necessary to consider more complex systems than SLs presented here. If graphene structure's conductivity is taken as depending on the wave vector, the nonlocal effects arise for near-field physics (such as plasmonics) and for the optical properties (far-field spectroscopy). In particular, the usage of the graphene plasmonics to probe nonlocal effects within the metal thin film was recently proposed for the graphene/hBN/metal system in Ref. \cite{Dias}.

The placement of a graphene sheet at a distance of a few nanometers away from a metal surface was experimentally studied recently \cite{Lund}. The near-field imaging experiments provided an evidence for the existence of three types of nonlocal effects in the massless Dirac electron liquid: the single-particle velocity matching, the interaction-enhanced Fermi velocity, and the interaction-reduced compressibility.

\section{\label{s5}Dispersion relation for SPP's}

Turning to SPPs in the planar graphene SLs, we are starting from the macroscopic Maxwell's equations
\begin{equation}\label{28}
\begin{split}
\DIV{\bf D}&=4\pi\rho_f,\hspace{0.65cm}\DIV{\bf B}=0,\\
\ROT{\bf E}&=-\frac{1}{c}\frac{\partial{\bf B}}{\partial t},\hspace{0.25cm}\ROT{\bf H}=\frac{4\pi}{c}{\bf j}_f+\frac{1}{c}\frac{\partial{\bf D}}{\partial t},
\end{split}
\end{equation}
where ${\bf D}=\varepsilon_\ell{\bf E}$ and ${\bf B}=\mu_\ell{\bf H}$ are the vectors of electrical and magnetic induction, related to the electric ${\bf E}$ and magnetic ${\bf H}$ field strengths, respectively, via dc permittivity $\varepsilon_\ell$ and dc permeability $\mu_\ell$ of media surrounding the system, $\ell=1,\,2$ (for the sake of generality, we shall not yet assume $\mu_\ell=1$), $\rho_f$ and ${\bf j}_f$ are the charge density and the current density, respectively. We take $\varepsilon_\ell$ and $\mu_\ell$ in the static limit, since, as we will see below, we are dealing with small frequencies ($\omega<10$ meV).

\begin{figure}[t!]
\begin{center}
\includegraphics[width=0.48\textwidth]{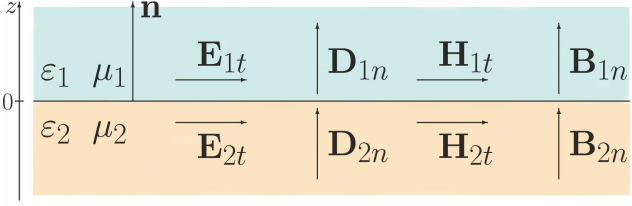}
\caption{\label{f5} A configuration of the tangential and perpendicular components of fields at the interface $z=0$ where the planar graphene SL is located.}
\end{center}
\end{figure}

Here, $xy$ plane lies in the surface of the system. Then, $\rho_f=\rho_s\delta(z)$ and ${\bf j}_f={\bf j}_s\delta(z)$ with the surface charge density $\rho_s$ and the surface current density ${\bf j}_s$. We have also the material equation (in the quasi-2D case)
\begin{equation}\label{29}
{\bf j}_s=\sigma_{xx}E_x\textbf{e}_x+\sigma_{yy}E_y\textbf{e}_y,
\end{equation}
where $\sigma_{xx}$ and $\sigma_{yy}$ are the diagonal components of the optical conductivity tensor of the system. The tangential component of the electric field strength ${\bf E}_t=(E_x,\,E_y,\,0)$ lies in the $xy$ plane (see Fig. \ref{f5}). In the quasi-1D case, we have to modify the relation (\ref{29}) because of a different dimensionality of the surface current $I_s=\sigma E_y$ (as the current in the system along the $y$ direction in one 1D element) and the surface current density ${\bf j}_s$. We should analogously introduce the value ${\bf j}_s=(0,\,j_s,\,0)$ with $j_s=I_s/d$ where $d$ is a characteristic dimension of the system along the $x$ direction (the SL period).

We recall that a graphene sheet and planar systems of monomolecular thickness based on it don't have their own dc permittivity and dc permeability, and, from the point of view of the electrodynamics of continuous media, they are actually an infinitely thin conductive layer between two dielectric media. Moreover, these media can be considered infinitely thick, occupying half spaces under the graphene system ($z<0$) and above it ($z>0$).

We direct the normal ${\bf n}$ to the interface of these media along the $z$ axis. We denote the medium in the half space $z>0$ (this can be air or vacuum) as the medium with the number $\ell=1$, and the medium in the half space $z<0$ (most likely it is the substrate material) as the medium with the number $\ell=2$, i.e., the normal is directed from medium 2 to medium 1 (see Fig. \ref{f5}).

The boundary conditions at the interface $z=0$ are
\begin{equation}\label{30}
\begin{split}
{\bf E}_{1t}&={\bf E}_{2t},\hspace{0.25cm}{\bf H}_{1t}-{\bf H}_{2t}=\frac{4\pi}{c}\left[{\bf j}_s{\bf n}\right],\\
D_{1n}-D_{2n}&=4\pi\rho_s,\hspace{0.25cm}B_{1n}=B_{2n}.
\end{split}
\end{equation}

\begin{figure*}[t!]
\begin{center}
\includegraphics[width=0.42\textwidth]{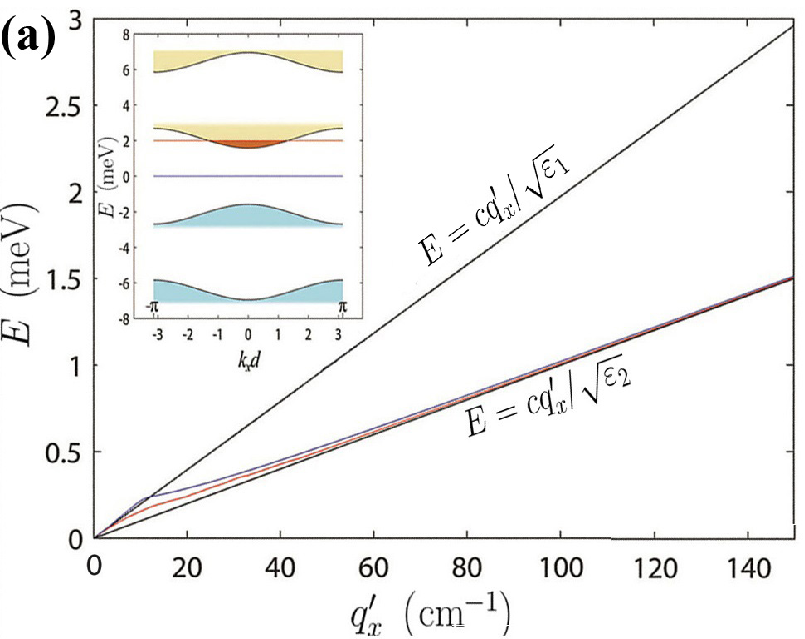}\hspace{0.5cm}\includegraphics[width=0.46\textwidth]{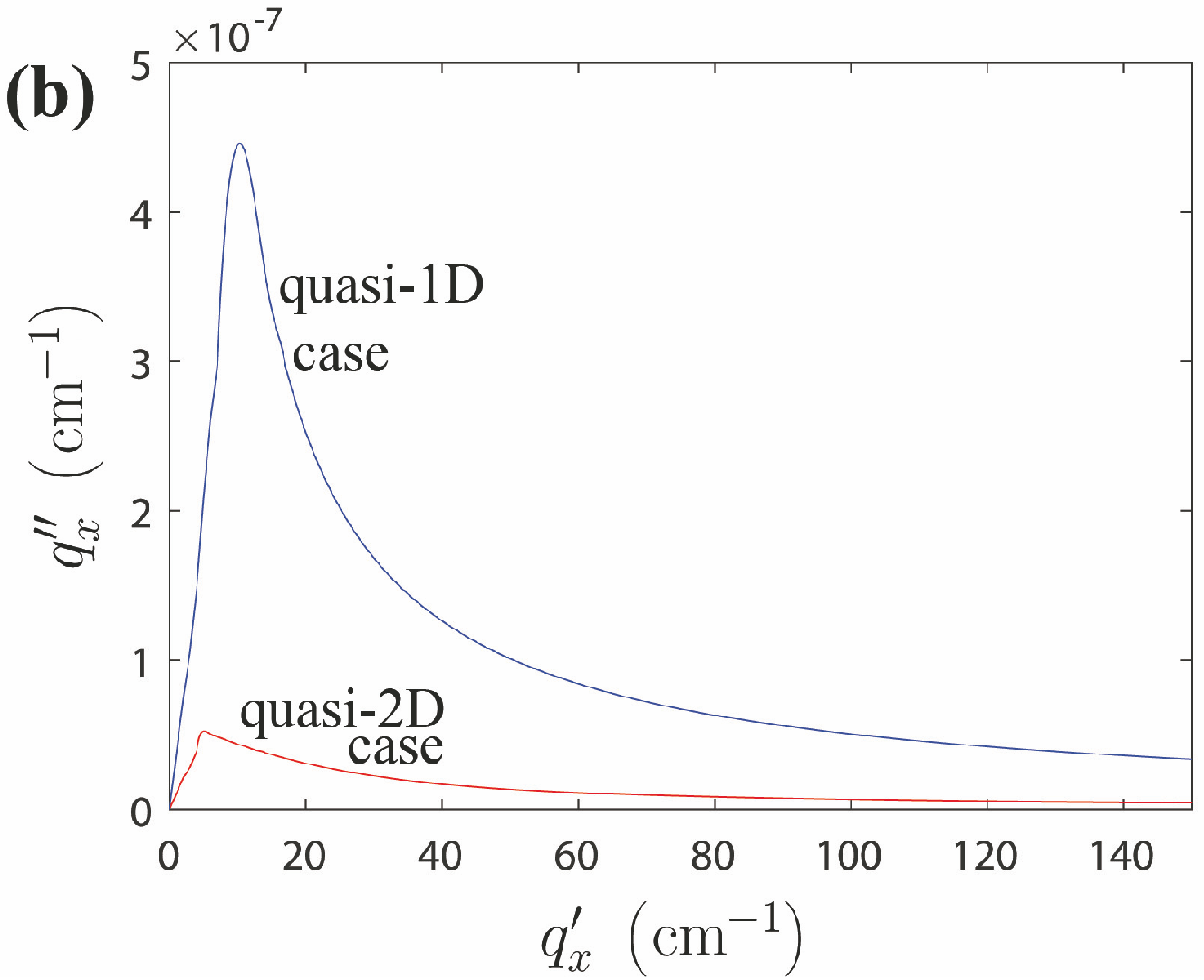}
\caption{\label{f6} (a) The dispersion dependence of TE$_0$ mode for the case of the Fermi level between the lower electron and the upper hole minibands (the blue curve) and the case of the Fermi level in the lower electron miniband (the red curve). The position of the Fermi level with respect to minibands is shown in the inset by lines of different color. (b)~The dependence of the imaginary part of the wave vector $q^{\prime\prime}_x$ on its real part $q^\prime_x$.}
\end{center}
\end{figure*}

Since the planar graphene SL is a 2D system, it enters in the calculation of the dispersion relation for SPPs only through the boundary conditions Eqs. (\ref{30}) together with the material Eq. (\ref{29}). Consequently, we need to know only its optical conductivity $\sigma$.

Considering the absence of free volume currents and charges, we are looking for a solution to the system of Maxwell's equations in each medium in the form
\begin{equation}\label{31}
\begin{split}
{\bf E}_\ell({\bf r},\,t)&={\bf E}_\ell(x,\,z)e^{i({\bf q}{\boldsymbol\varrho}-\omega t)},\\
{\bf H}_\ell({\bf r},\,t)&={\bf H}_\ell(x,\,z)e^{i({\bf q}{\boldsymbol\varrho}-\omega t)},
\end{split}
\end{equation}
where ${\bf q}=(q_x,\,q_y,\,0)$ and ${\boldsymbol\varrho}=(x,\,y,\,0)$ are 2D vectors in the $xy$ plane (${\bf q}$ is the wave vector), and vectors ${\bf E}_\ell(x,\,z)$ and ${\bf H}_\ell(x,\,z)$ are defined as periodical functions of $x$ with the period $d$ which coincides with the SL period, ${\bf E}_\ell(x,\,z)={\bf E}_\ell f(x,\,z)$ and ${\bf H}_\ell(x,\,z)={\bf H}_\ell f(x,\,z)$ and $f(x+\lambda d,\,z)=f(x,\,z)$ for any $z$ and $\lambda\in\mathbb{Z}$.

The remaining fields are expressed by the relations
\begin{equation}\label{32}
\begin{split}
{\bf D}_\ell({\bf r},\,t)&=\varepsilon_\ell{\bf E}_\ell({\bf r},\,t),\\
{\bf B}_\ell({\bf r},\,t)&=\mu_\ell{\bf H}_\ell({\bf r},\,t).
\end{split}
\end{equation}

So, we can write the function $f(x,\,z)$ as the Fourier-Floquet series
\begin{equation*}
f(x,\,z)=\sum\limits_{\nu=-\infty}^\infty f_\nu e^{2\pi i\nu x/d}e^{-\kappa_{\ell\nu}|z|},
\end{equation*}
where $f_\nu$ are numbers determined by the Fourier integral with the function $f(x,\,z)$; $\kappa_{\ell\nu}$ are wavenumbers which define an exponential decay of the fields in each medium. The action of the derivative with respect to $x$ on the fields reduces to multiplying the terms of the series by $iq_{x\nu}=iq_x+i\nu G$, where $G=2\pi/d$ is the 1D reciprocal lattice wave vector.

After substitution the fields Eqs. (\ref{31}) and (\ref{32}) into Eqs. (\ref{28}), we obtain a system of linear equations, the compatibility condition of which gives the relation
\begin{equation}\label{33}
\kappa^2_{\ell\nu}=q^2_{x\nu}+q^2_y-\varepsilon_\ell\mu_\ell\frac{\omega^2}{c^2}.
\end{equation}

After simple calculations, we have the following dispersion relation for SPPs:
\begin{equation}\label{34}
\begin{split}
\frac{q^2_{x\nu}q^2_y}{\varkappa^2_\nu}-&\left(\widetilde{\varkappa}_\nu-\frac{q^2_{x\nu}}{\varkappa_\nu}-\frac{4\pi i\omega}{c^2}\sigma_{xx}\right)\\
&\times\left(\widetilde{\varkappa}_\nu-\frac{q^2_y}{\varkappa_\nu}-\frac{4\pi i\omega}{c^2}\sigma_{yy}\right)=0,
\end{split}
\end{equation}
where $\varkappa^{-1}_\nu=\left(\mu_1\kappa_{1\nu}\right)^{-1}+\left(\mu_2\kappa_{2\nu}\right)^{-1}$ and $\widetilde{\varkappa}_\nu=\kappa_{1\nu}/\mu_1+\kappa_{2\nu}/\mu_2$.

Now, we consider two special cases.

(a) \emph{The wave vector is directed along the $x$ axis, $q_x\neq0$ and $q_y=0$}. If $E_x=0$ and $E_y\neq0$ (as is easily seen, also $E_{\ell z}=0$), we have the dispersion relation for $\nu$th transverse electric (TE$_\nu$) mode of SPPs \cite{BFPV}:
\begin{equation}\label{35}
\frac{\kappa_{1\nu}}{\mu_1}+\frac{\kappa_{2\nu}}{\mu_2}-\frac{4\pi i\omega}{c^2}\sigma_{yy}=0.
\end{equation}
If $E_x\neq0$ and $E_y=0$ (as is easily seen, also $H_{\ell x}=H_{\ell z}=0$ and $H_{\ell y}\neq0$), we have the dispersion relation for $\nu$th transverse magnetic (TM$_\nu$) mode of SPPs \cite{BFPV}:
\begin{equation}\label{36}
\frac{\varepsilon_1}{\kappa_{1\nu}}+\frac{\varepsilon_2}{\kappa_{2\nu}}+\frac{4\pi i}{\omega}\sigma_{xx}=0.
\end{equation}
It should be emphasized that the relations Eqs. (\ref{35}) and (\ref{36}) hold for any $\nu$. The spectrum of TM modes exist only in the quasi-2D case because there is no transfer of charge carriers along the $x$ direction in the quasi-1D case (formally, $\sigma_{xx}\rightarrow0$).

\textbf{(b)} \emph{The wave vector is directed along the $y$ axis, $q_x=0$ and $q_y\neq0$, and $\nu=0$}. The system of equations describing SPPs is obtained from the system of equations considered for the above case by the substitution $x\rightleftarrows y$. So, if $E_x=0$ and $E_y\neq0$, we have TM$_0$ mode of SPPs [the dispersion relation is Eq. (\ref{35}) with $\nu=0$] and, if $E_x\neq0$ and $E_y=0$, we have TE$_0$ mode of SPPs [the dispersion relation is Eq. (\ref{34}) with $\nu=0$].

Let us demonstrate the difference between cases of altered positions of the Fermi level on an example of a TE$_0$ mode propagating along the $x$ axis. We consider SL with gapped graphene creating by deposition of CrO$_3$ molecules with the half width of the band gap $\Delta_0=60$ meV \cite{Zan}. For simplicity, we took $V_0=0$. The width of gapless graphene stripes is $d_\text{I}=403.36$ nm (1640 unit cells) and the width of gapped graphene stripes is $d_\text{II}=14.76$ nm (60 unit cells). The substrate is the silicon dioxide with the dielectric constant $\varepsilon_2=3.9$ (above the system is vacuum or air with $\varepsilon_1=1$ and $\mu_1=\mu_2=1$). We assume that the Fermi level falls within the minigap between the lower electron and the upper hole minibands (there is no the Drude contribution to the optical conductivity of the system, because free charge carries are absent), and then its position can be changed by the electric field effect and it is located within the lower electron miniband (the upper hole miniband). We took the inverse relaxation times $\gamma=24$ meV and $\Gamma=1$ meV to obtain $\IM\sigma_{yy}<0$, which is a necessary condition for the existence of a solution to the dispersion Eq. (\ref{35}) (it~is clear that the intraminiband relaxation time must be much smaller than the interminiband relaxation time). The results for the dispersion dependence of the TE$_0$ mode at small wave vectors are presented in Fig. \ref{f6}.

The blue curve shows the dispersion of the TE$_0$ mode for the case of the Fermi level between the lower electron and the upper hole minibands. It starts above the upper light cone. This is a consequence of the reduced optical conductivity (without the Drude contribution). The attenuation of SPPs is also enhanced: The imaginary part of the wave vector $q^{\prime\prime}_x$ is almost an order of magnitude large than in the quasi-2D case. The position of the peak of the blue curve for $q^{\prime\prime}_x$ corresponds to the intersection of the blue curve for $E$ with the upper light cone, and the peak of the red curve for $q^{\prime\prime}_x$ corresponds to a sharp deviation of the dispersion curve for the quasi-2D case from the upper light cone toward the lower one.

\begin{figure}[h!]
\begin{center}
\includegraphics[width=0.48\textwidth]{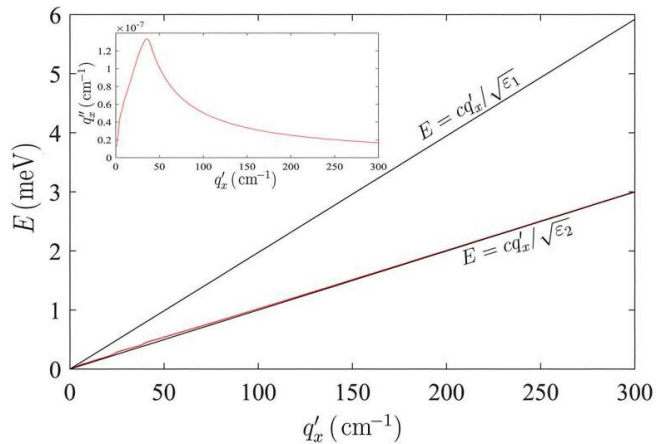}
\caption{\label{f7} The dispersion dependence of TM$_0$ mode for the case of the Fermi level in the lower electron miniband. The dependence of the imaginary part of the wave vector $q^{\prime\prime}_x$ on its real part $q^\prime_x$ is shown in the inset.}
\end{center}
\end{figure}

For comparison, we present in Fig. \ref{f7} the results for the dispersion dependence of the TM$_0$ mode propagating along the $x$ axis. At the same time, we took into account that the necessary condition for the reality of the solutions of Eq. (\ref{36}) is $\IM\sigma_{xx}>0$. To ensure this condition, we took the values $\gamma=1.5$ meV and $\Gamma=1$ meV. Second, as mentioned above, we have the TM modes spectrum with $q_x\neq0$ and $q_y=0$ only in the quasi-2D case. Therefore, we performed calculations when the Fermi level falls into one of the minibands (we chose the lower electron miniband and $E_F=2$ meV).

A distinctive feature of the TM$_0$ mode in comparison with the TE$_0$ mode is that its dispersion curve is rapidly pressed to the lower light cone. This is a consequence of the reduced optical conductivity $\sigma_{xx}$ due to the presence of a small parameter $v_\perp/v_\parallel$. The curve $q^{\prime\prime}_x(q^\prime_x)$ also has a peak, as for the TE$_0$ mode. We emphasize that the influence of SLs on the SPPs spectrum consists in the fact that two types of motion of charge carriers are possible.

\section{\label{s6}The estimation of losses at excitation of SPP's}

SPP excitation methods are widely discussed in the scientific literature (see, e.g., Ref. \cite{BFPV} and references therein). The apex of an illuminated nanoscale tip can be effectively applied for SPP excitation \cite{Fei}. Such a realistic experimental situation corresponds to the local spatial excitation of SPPs (they have a complex wave vector and real frequency, as used here).

The estimation of losses due to SPP excitation is in many ways similar to the estimation of the absorption of external electromagnetic radiation on the excitation of plasmons in SLs \cite{Rat1}. We use the well-known formula~\cite{LL}
\begin{equation}\label{37}
Q=\frac{1}{2}\RE\left(\sigma\mathbf{E}\widetilde{\mathbf{E}}^*\right),
\end{equation}
where $\sigma$ is the optical conductivity of the system, $\mathbf{E}$ is the SPP electric field, and $\widetilde{\mathbf{E}}$ is the electric field of the external electromagnetic wave.

Following Ref. \cite{Chap}, we can similarly derive \cite{Rat1}

(${\boldsymbol\alpha}$) \emph{in the quasi-1D case},
\begin{equation}\label{38}
\frac{Q^\text{(1D)}}{|\mathbf{E}_0|^2}\simeq\frac{\omega^2\gamma^2\sigma^\text{intra}_0}{(\omega^2-\omega^2_q)^2+\omega^2\gamma^2}+
\frac{\omega^2\Gamma^2\sigma^\text{inter}_0}{(\omega^2-\omega^2_q)^2+\omega^2\Gamma^2},
\end{equation}
where $\sigma^\text{intra}_0$ and $\sigma^\text{inter}_0$ are the values of conductivity Eqs. (\ref{20}) and (\ref{25}), (\ref{26}) in the zero-frequency limit, respectively, $\omega_q=\omega(q)$ is SPP frequency corresponding to the wave vector $q$, $\mathbf{E}_0$ is the electric field amplitude, $\mathbf{E}_0=(0,\,E_0,\,0)$;

(${\boldsymbol\beta}$) \emph{in the quasi-2D case},
\begin{equation}\label{39}
\frac{Q^\text{(2D)}_\perp}{|\mathbf{E}_0|^2}\simeq\frac{\omega^2\gamma^2\sigma^\text{intra}_{xx0}}{(\omega^2-\omega^2_q)^2+\omega^2\gamma^2}+
\frac{\omega^2\Gamma^2\sigma^\text{inter}_{xx0}}{(\omega^2-\omega^2_q)^2+\omega^2\Gamma^2},
\end{equation}
where $\sigma^\text{intra}_{xx0}$ and $\sigma^\text{inter}_{xx0}$ are the values of conductivity Eqs. (\ref{21}) and (\ref{23}), (\ref{24}) in the zero-frequency limit, respectively, $\mathbf{E}_0=(E_0,\,0,\,0)$. For polarization $\mathbf{E}_0=(0,\,E_0,\,0)$, we have $Q^\text{(2D)}_\parallel$ with the conductivity $\sigma^\text{intra}_{yy0}$ and $\sigma^\text{inter}_{yy0}$.

\section{\label{s7}Conclusions}

We have considered here SPPs in the planar graphene SLs with 1D periodic modulation of the band gap and obtained the dispersion relation for them. In this paper, we have demonstrated the opportunity for the transformation of the SPP spectrum due to a change of the optical conductivity in the system. This change can be achieved owing to variations of the Fermi-level position by the electric field effect. At sufficiently enough narrow minibands and minigaps, the Fermi level can be easily shifted from a minigap to neighbour miniband. In the case when the Fermi level falls within the minigap, there is a quasi-1D motion of charge carriers (excluding the case of the minigap between the lower electron and the upper hole minibands when charge carriers are absent). In the case when the Fermi level falls within the miniband, there is a quasi-2D motion of charge carriers. Thus, there arises a kind of 1D/2D-crossover in behaviour of charge carriers. This causes a significant difference in the optical conductivity of SLs and the SPP spectrum becomes tunable.

Various promising materials are now considered as candidates for active tuning of SPPs, including graphene and its gap modifications. The application of these materials to nanoelectronics is currently particularly attractive for the development of \emph{planar technology} for integrated circuits of the new generation. We expect that the creation and experimental study of planar graphene heterostructures can play a key role in achieving this goal.

\begin{acknowledgments}
The author is grateful to S. G. Tikhodeev for the helpful discussion and valuable advice on this publication. The work was supported by the Foundation for the Advancement of Theoretical Physics and Mathematics BASIS (the general formulation of the problem) and by the Russian Science Foundation (Project No. 16-12-10538-$\Pi$, the calculation of the optical conductivity, Sec.~\ref{s4}).
\end{acknowledgments}

\nocite{*}
\bibliography{SPPinPGSL}

\end{document}